\documentclass[amsmath,amssymb,aps,prb,twocolumn,groupedaddress,showpacs,floatfix]{revtex4}
\usepackage{graphicx}
\begin{document}

\newcommand {\be}{\begin{equation}}
\newcommand {\ee}{\end{equation}}
\newcommand {\bea}{\begin{eqnarray}}
\newcommand {\eea}{\end{eqnarray}}
\newcommand {\nn}{\nonumber}

\title{Quantum Phase Transitions in Coupled Dimer Compounds} 

\author{Omid Nohadani$^{(1)}$}
\author{Stefan Wessel$^{(2)}$}
\author{Stephan Haas$^{(1)}$}

\affiliation{$^{(1)}$Department of Physics and Astronomy, University of Southern California, Los Angeles, CA 90089-0484}
\affiliation{$^{(2)}$Institut f\"ur Theoretische Physik III, Universit\"at Stuttgart, 70550 Stuttgart, Germany}

%\date{\today} 
\begin{abstract} 
We study the critical properties in cubic systems of antiferromagnetically coupled spin dimers
near magnetic-field induced quantum phase transitions.
The quantum critical points in the zero-temperature phase diagrams are determined from quantum Monte Carlo simulations.
Furthermore, scaling properties of the uniform magnetization and the staggered transverse magnetization
across the quantum phase transition in magnetic fields are calculated.
The critical exponents are derived from Ginzburg-Landau theory.
We find excellent agreement between the quantum Monte Carlo simulations and the analytical results.
\end{abstract}


\maketitle

\section{\label{int_sec}Introduction}
\begin{figure}[h]
\includegraphics[width=8cm]{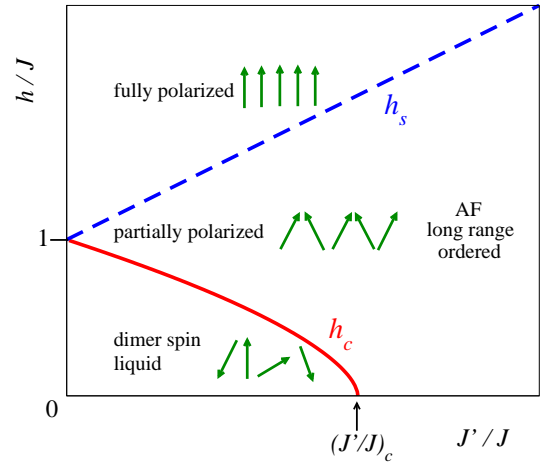}
\vspace{-2.2cm}
\caption{\label{phase_diagram}
Schematic zero-temperature phase diagram of a three-dimensional coupled dimer compound
 with intra-dimer couplings $J$ and inter-dimer couplings $J'$. 
$h$ denotes the strength of the external magnetic field.
}
\end{figure}

Recent improvements in high magnetic field technology have made detailed investigations of quantum  phenomena in strong magnetic fields possible. In particular, they allow for 
studies of quantum critical properties induced by high magnetic fields in weakly coupled spin dimer compounds, such as 
$\rm TlCuCl_3$ \cite{rsttktmg,tlcucl,rtokuokh}, $\rm KCuCl_3$ \cite{rsttktmg,kcucl},
$\rm BaCuSi_2O_6$ \cite{bacusio}, and $\rm Sr_2Cu(BO_3)_2$ \cite{srcubo}.
The ground state of these materials consists of local spin singlets. 
If a magnetic field is applied that exceeds their singlet-triplet excitation gap, 
they undergo a transition 
into a magnetically ordered state.
This quantum phase transition into a regime with transverse antiferromagnetic (AF) order
can be described as a Bose-Einstein condensation (BEC) of triplet excitations, which behave as bosonic 
quasiparticles, called triplons.
In the corresponding Bose-Hubbard model, this is analogous to the transition from the Mott-insulating phase to the superfluid condensate,
where the magnetic field translates into the chemical potential.
At very high magnetic fields,
there is a saturation threshold, beyond which the spins fully align
along the field direction, and the transverse antiferromagnetic order is destroyed.
A schematic zero-temperature phase diagram is shown in Fig.\ref{phase_diagram}.
Because these compounds are three-dimensional,
field-induced transverse antiferromagnetic order persists up to a 
finite transition temperature, $T_c(h)$, between the upper and lower critical field ($h_c\le h\le h_s$).
Approaching the critical field from the partially polarized phase, $h\rightarrow h_c$,
the critical temperature is expected to vanish as 
$T_c (h) \propto |h - h_{c}|^{1/\alpha}$, with a universal power-law exponent that is predicted to be 
$\alpha=3/2$.~\cite{sachdevbook,kawashima_qcp}
While early experimental \cite{rtokuokh} and numerical \cite{wessel} studies reported non-universal, i.e. coupling dependent, values of $\alpha>2$, it was recently shown in Ref. \onlinecite{universality} that 
careful fitting of both experimental and numerical data indicates an effective exponent $\alpha(h)>3/2$, which approaches $3/2$
as $h\rightarrow h_c$. Motivated by this observation, Shindo and Tanaka\cite{shindo_tanaka} performed a 
fit of specific heat data in $\rm TlCuCl_3$ in the close vicinity of the lower critical field, yielding an exponent of $\alpha = 1.67\pm0.07$. This value is closer to the
universal value of $\alpha$ found in Ref. \onlinecite{universality} than previous
experimental reports.

\begin{figure}[h]
\includegraphics[width=9cm]{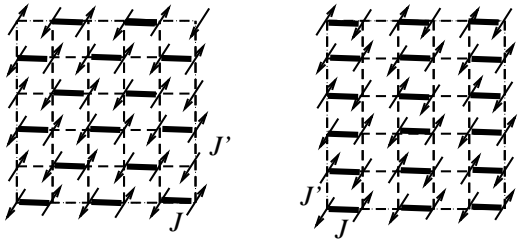}
\vspace{-6.0cm}
\caption{Layers of coupled dimers, with staggered (left panel) and aligned (right panel) arrangements of dimers. The
dimer bonds ($J$) are denoted by solid bars, whereas the inter-dimer bonds ($J'$) are denoted by dashed lines. In the 
three-dimensional crystal, these layers are coupled by interlayer couplings of the same strength as $J'$.
The arrangement of the dimers in neighboring layers is staggered (left panel), or aligned (right panel), respectively.}
\label{structures} 
\end{figure}

While numerical and experimental results on critical properties of
the field-induced ordering transition agree well with
analytical results at finite temperatures\cite{bec,kawashima_qcp}, there are no
comparable scaling predictions for the zero-temperature quantum phase transitions.
In this paper, we present a numerical analysis of critical properties
in cubic systems of coupled dimers. In particular,
zero-temperature scaling properties of the uniform and staggered magnetization
in the partially polarized phase are studied, based
on large-scale quantum Monte Carlo (QMC) simulations.
These systems, upon increasing the inter-dimer coupling, enter an antiferromagnetically ordered state at a quantum critical point
$(J'/J)_c$.~\cite{sachdevbook} 
Magnetic exchange constants are known to depend sensitively on the
distance between the magnetic sites. If their relative magnitudes are   
altered by the application of external pressure, such magnetic ordering transitions can be induced, as recently observed for $\rm TlCuCl_3$.
\cite{tanaka_pressure,tanaka_pressure2}
In this numerical study, the quantum critical point is determined for the
structures shown in Fig.~\ref{structures}. Furthermore,
the quantum criticality induced by an applied magnetic field is studied. 
In particular, the scaling behavior of the uniform magnetization and 
the antiferromagnetic order parameter upon entering the partially polarized phase is determined. The associated critical exponents
can be accessed experimentally and thus allow for a quantitative comparison.

The paper is organized as follows:
in the next section, the model used in the numerical simulations is defined and details about the quantum Monte 
Carlo method are presented. In Sec. \ref{phasediag_sec}, the zero-temperature  phase diagram of the systems is discussed, and the 
quantum critical points at zero magnetic field are determined.
A detailed scaling analysis of the magnetic properties
at the quantum phase transitions in a magnetic field is presented in Sec. \ref{qpt_sec}. Finally, conclusions are given in Sec.\ref{sum_sec}.
The Appendix contains
a phenomenological derivation of the mean-field exponents, observed in Sec. \ref{qpt_sec}, within Ginzburg-Landau theory.

\section{\label{num_sec}Model and Method}

We consider the spin-1/2 Heisenberg antiferromagnet on the lattice structures shown in Fig.~\ref{structures}. The Hamiltonian is given by
\bea
H = \sum_{\langle i,j \rangle} J_{ij} {\bf S}_i \cdot {\bf S }_j - h \sum_i S^z_i ,
\eea
where the ${\bf S}_i$ denote localized spin-1/2 moments, and 
$J_{ij}$ indicates the coupling constant between sites $i$ and $j$, which takes values $J$ for the dimer, and J' 
for the inter-dimer couplings. Furthermore, $h$ denotes the applied magnetic field. 

Since the lattice structures that are considered here, are bipartite
(c.f. Fig.~\ref{structures}), antiferromagnetism in these systems is  
not frustrated,
so that their properties can be studied 
using large-scale QMC, without a sign-problem~\cite{sgn}.
Here, we use the stochastic series expansion QMC
method~\cite{sse,sandvik_green}
with directed loop updates~\cite{directed_loop,alet}.
This update scheme results in 
higher sampling efficiency than other methods based
on the conventional operator-loop 
update.  Close to criticality, 
autocorrelation times are reduced by up to an order of magnitude.
This allows detailed simulations of ground state properties on 
clusters with up to 10$\;\!$000 sites, even in the presence of large magnetic fields.
Furthermore, off-diagonal observables such as the transverse magnetic structure factor 
can be measured efficiently  during the directed loop construction.\cite{troyer,alet}
Therefore, the 
order parameter in the partially polarized phase, i.e. the staggered transverse magnetization perpendicular to the magnetic field direction, can be calculated by using
\begin{equation}
\label{perp_stagg_mag}
m_{\rm s}^{\perp} = \sqrt{\frac{S^\perp_{\rm s}}{L^3}}.
\end{equation}
Here, $L$ denotes the linear system size and $S^\perp_{\rm s}$ the 
transverse staggered structure factor,
\begin{equation}
\label{stagg_structure_factor}
S^\perp_{\rm s} = \frac{1}{L^3}\sum_{\langle i,j \rangle} (-1)^{i+j}
 \langle S^{x}_i S^{x}_j \rangle .
\end{equation}
These observables have been shown to saturate in QMC simulations at temperatures below $T=J'/2L$, which is used throughout this study to ensure zero-temperature behavior.

\section{\label{phasediag_sec}Zero-Temperature Phase Diagram}

The zero-temperature  
phase diagram of coupled dimer systems in a magnetic field
can be obtained using the bond-operator mean-field theory~\cite{matsumoto,matsumoto_pressure}. 
A schematic phase diagram is shown in Fig.~1.
It consists of three
phases: (i) at low fields and small coupling ratios $J'/J$, the system
is in a magnetically disordered phase, i.e. a dimer spin liquid; (ii) at intermediate fields
and/or sufficiently large values of  $J'/J$, the ground state is partially spin
polarized and has an antiferromagnetic long-range order transverse to the magnetic field direction; and
(iii) at large fields, $h>h_s=J+5J'$ all spins are fully polarized. 
These phase separations occur in both dimer arrangements, shown
in Fig.~\ref{structures}.

While bond-operator theory provides a reliable description of the phase diagram~\cite{universality},
a more precise estimate of the critical inter-dimer coupling strength, $(J'/J)_c$, is required for the
study of critical properties in finite magnetic fields, presented below.
In order to determine this zero-field quantum critical point, 
we perform a finite-size scaling analysis of 
the staggered magnetization obtained from QMC simulations for various system sizes. Defining the dimensionless coupling ratio  $g=J'/J$, 
the relevant finite-size scaling is obtained as follows.
The correlation length $\xi$ diverges near the quantum critical point $g_c$ as $\xi \propto |g-g_c|^{- \nu}$. The correlation time $\tau_c$
during which fluctuations relax and decay (equilibration),
is related to the correlation length $\xi$ via
$\tau_c \propto \xi^z \propto |g-g_c|^{- \nu z}$,
with the dynamical critical exponent $z$.\cite{vojta_qpt}
In the vicinity of the critical point and at zero-field,
the staggered magnetization $m_{\rm s}^\perp = m_{\rm s}$ for $g>g_c$ scales as
\be
\label{stag_mag_scale}
m_{\rm s} \propto (g-g_c)^\beta ,
\ee
defining an exponent $\beta$.\cite{troyer_exponent}
In general, this implies a finite-size scaling relation of the 
order parameter at the quantum critical point, $g=g_c$:
\be
\label{stag_mag_L_scaling}
m_{\rm s} \propto L^{-\beta/\nu}.
\ee
Since these systems are explicitly dimerized, the Berry-phase contributions to the path-integral cancel.~\cite{science}
Therefore, this QPT
belongs to the universality class of the classical $O(3)$ Heisenberg model in $3+1$ dimensions, with the dynamical critical exponent $z=1$.
Since the effective classical model is at the upper critical dimension ($d_c=4$), 
the critical exponents $\nu$ and $\eta$ take on mean-field values
$\nu=1/2$, $\eta=0$, and the above scaling laws hold up to logarithmic corrections.\cite{sachdevbook}

\begin{figure}[h]
\includegraphics[width=8cm]{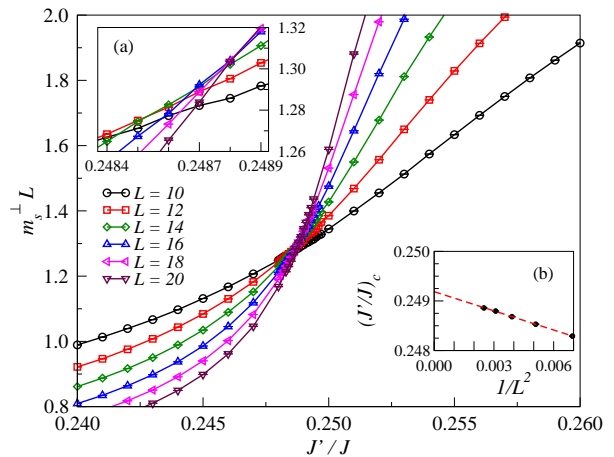}
\vspace{-0.3cm}
\caption{\label{crit_coupl}
Scaling plot of the zero-temperature staggered magnetization in the aligned arrangement of Fig.~2~(b) obtained 
from quantum Monte Carlo simulations using systems of linear sizes $L=10$ to $20$. $J$ denotes the intra-dimer coupling. 
At the critical inter-dimer coupling $(J'/J)_c$, the
different curves intersect each other (main panel). Inset (a) shows a magnification
of the intersection region. Inset (b) exhibits the corrections to the mean-field scaling behavior, arising from logarithmic corrections in the upper critical dimension.
The statistical error bars fall within the symbol size.}
\end{figure}

A scaling plot of $m_{\rm s}^\perp L$ is shown in Fig.~\ref{crit_coupl}, using data from system sizes $L=10$ to $20$.
On the scale of the main part of Fig.~\ref{crit_coupl}, a common intersection point of the finite-size data appears to exist.
However, on a smaller scale, the insets of Fig.~\ref{crit_coupl} exhibit the logarithmic corrections to Eq.(\ref{stag_mag_L_scaling}).
With increasing system size, we observe that the
intersections for neighboring system sizes move towards increasing values of $(J'/J)$.
As seen 
in inset (b) of Fig.~\ref{crit_coupl}, the systematic increase of the crossing points scales well as a function of $1/L^2$, which allows 
us to extrapolate the thermodynamic limit value of
$(J'/J)_c = 0.2492\pm0.0002$ for the quantum critical point.
These logarithmic corrections are best accessible when system-sizes
of different orders of magnitude are compared.
However, because of computational restrictions, we follow the approach described above.

Performing a similar analysis for the staggered configuration of dimers in Fig.~\ref{structures}(a),
we find the  quantum critical point at $(J'/J)_c =0.2170\pm0.0002$.
This lower critical inter-dimer coupling for the staggered configuration indicates
a reduced tendency towards the formation of dimer singlets for this arrangement of dimers, as compared
to the aligned dimer configuration.
\begin{figure}[h]
\includegraphics[width=8.5cm,height=7cm]{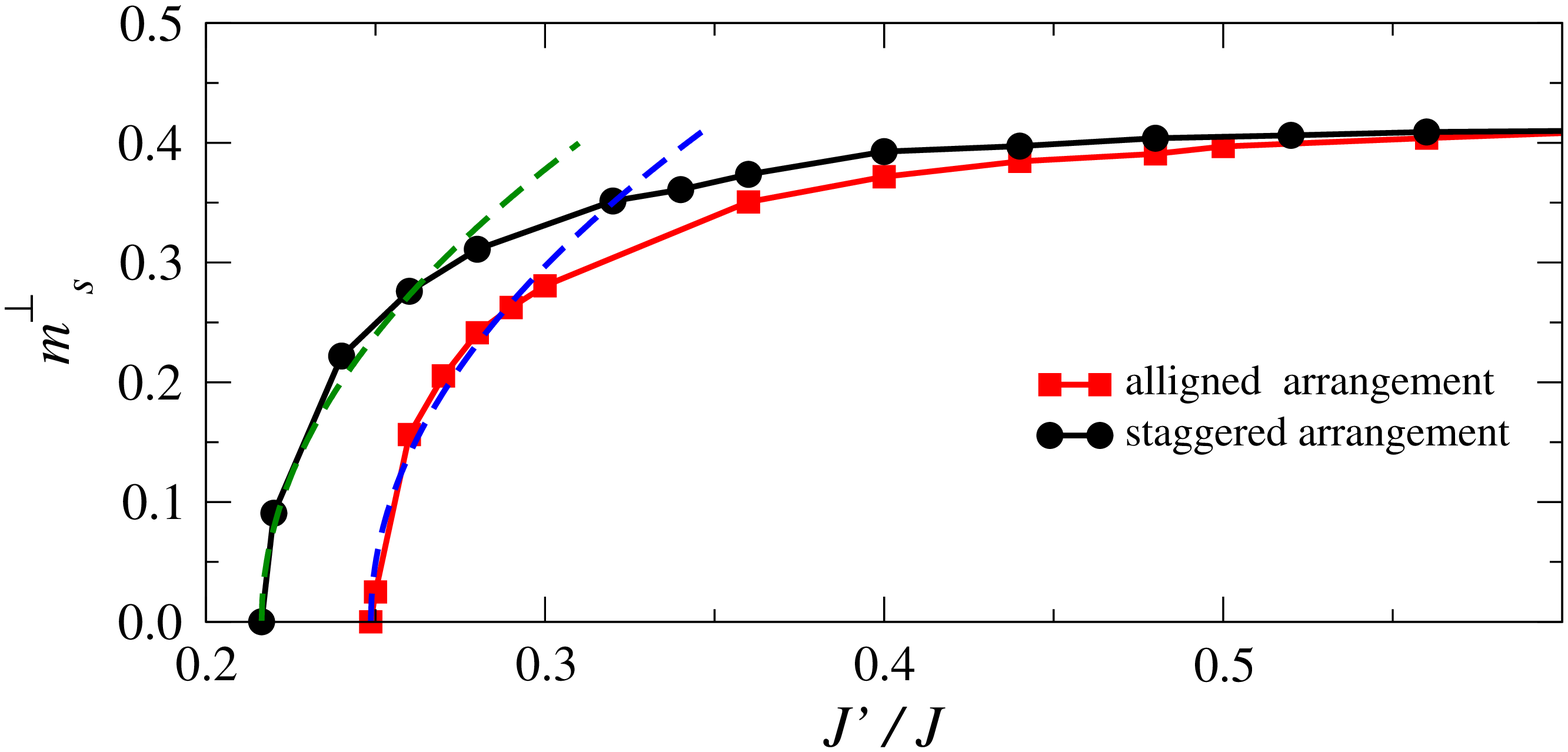}
\vspace{-3cm}
\caption{\label{m_j} 
Staggered magnetization $m_s$ of coupled dimer arrays as a function of the inter-dimer coupling strength $J'$,
for the two different dimer arrangements of Fig. 2. Dashed lines show the mean-field scaling behavior with
exponent $\beta=\frac{1}{2}$ close to the quantum critical point.
} 
\end{figure}

Beyond the critical inter-dimer coupling, the staggered magnetization increases with a mean-field exponent $\beta=1/2$, as in Eq.~(\ref{stag_mag_scale}). The QMC data of $m_s$, shown in Fig.~\ref{m_j}, are consistent with this scaling behavior.
However, due to restricted system sizes, we are not able to extract 
the logarithmic corrections to the scaling law.

\section{\label{qpt_sec}Field-Induced Quantum Phase Transition}

Having determined the quantum critical points of coupled-dimer arrays, we proceed to study the 
effects of a magnetic field in the different regimes. In particular, we determine the scaling behavior of the uniform magnetization
and the order parameter upon entering the partially polarized region. The corresponding scaling exponents
are accessible experimentally by direct measurement of the magnetization, and by neutron scattering.
Thus, the numerical results can be compared with analytical predictions and with measurements on the materials mentioned above.
Here, simulations for the aligned configuration of dimers, shown in Fig.~\ref{structures}(b), are presented.
For the staggered  configuration of dimers, the same scaling exponents are obtained. This is due to the underlying universality.

\subsection{\label{mu_sec}Scaling of the Uniform Magnetization}

First, we discuss the behavior of the uniform magnetization $m_u(h)$ as a function of the applied magnetic field. 
In the dimer spin liquid phase, i.e. for coupling ratios smaller than
$(J'/J)_c$, a finite magnetic excitation gap $\Delta$ separates the ground state
singlet and the lowest triplet state. Thus, a finite magnetic field $h_{c}=\Delta$ is required to close this gap, and to induce
a finite uniform magnetization.
The Ginzburg-Landau approach presented in Appendix A predicts $m_u$ to increase linearly, $m_u\propto (h-h_{c})$ 
for $h>h_{c}=\Delta$ and for $J'<J'_c$.
When the inter-dimer coupling is increased, the gap $\Delta$ becomes smaller, until it vanishes at the critical
inter-dimer coupling $J'_c$. At the quantum critical point, $J'=J'_c$, the Ginzburg-Landau approach
predicts the uniform magnetization to scale as $m_u\propto h^3.$ 
For larger values of $J' > J'_c$, the excitation
gap remains zero, and the finite uniform susceptibility $\chi_u$ results in a linear response $m_u=\chi_u h$ of the uniform 
magnetization in the N\'eel-ordered regime beyond $J'_c$.

\begin{figure}[h]
\includegraphics[width=8.5cm,height=7cm]{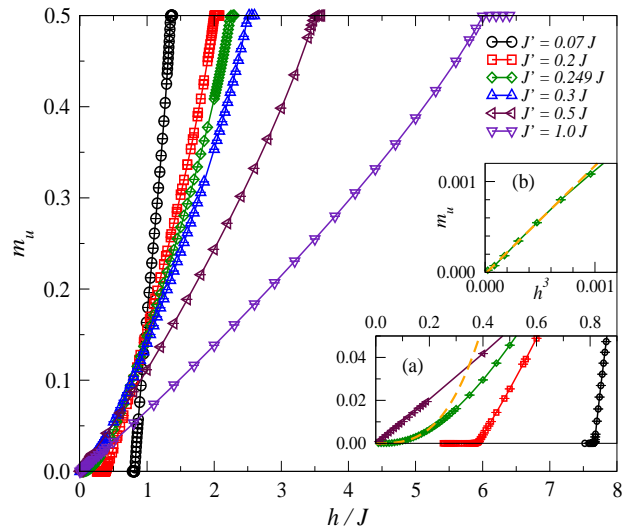}
\vspace{-0.4cm}
\caption{\label{umag} 
Zero-temperature uniform magnetization of aligned dimer arrays for different inter-dimer couplings as a function
of the magnetic field $h$. Inset (a) focuses on the low-field region, $h/J<1$. 
The scaling  $m_u\propto h^3$ at $J'=J'_c$ is demonstrated in inset (b).
}
\end{figure}

Let us now compare these predictions with the QMC results obtained from simulations 
of systems with linear sizes up to $L=30$. For such large system sizes, we did not detect finite-size effects in the uniform 
magnetization. In Fig.~\ref{umag}, results for various 
values of the inter-dimer coupling strength are presented.
The main part of  Fig.~\ref{umag} shows the uniform magnetization over the full range of magnetic field strengths, up to 
the saturation field $h_s=J+5J'$. Inset (a) of Fig.~\ref{umag} focuses on the region close to the critical field $h_c$ for various
inter-dimer couplings $J'$.
The linear scaling in $h$, which is predicted on both sides of the quantum critical point, is clearly observed in this small-$m_u$ region.
In contrast, at the quantum critical point, $J' = J'_c$, the uniform magnetization increases non-linearly with $h$, as shown in  Fig.~\ref{umag}~(a).
Indeed, we observe a scaling $m_u\propto h^3$, presented in Fig.~\ref{umag}~(b).
This is expected from the Ginzburg-Landau theory
(c.f. Appendix A) and the bond-operator mean-field theory\cite{matsumoto_pressure}.
Using QMC, these scaling exponents of the uniform magnetization are verified.

Finally, we note that the field dependence of $m_u$ for the entire region between $h_{c}$ and $h_{s}$ is linear for weakly coupled
dimers (see e.g. Fig.~\ref{umag} for $J'=0.07J$), consistent with
bond-operator mean-field theory~\cite{matsumoto_pressure}. 
For larger values of $J'$, $m_u$ shows deviations from this linear behavior in high magnetic fields. 
This non-linear behavior can be accounted for within bond-operator
theory by 
including the contributions of higher-energy triplet modes to the ground state~\cite{matsumoto_pressure}.

\subsection{\label{ms_sec}Scaling of the Order Parameter}

Next, we discuss the scaling properties of the order parameter in the partially polarized phase, 
i.e. the staggered transverse magnetization perpendicular to the 
magnetic field direction, $m_s^{\perp}(h)$, as a function of the applied magnetic field $h$.
Before presenting numerical data, the expectations from the Ginzburg-Landau theory from Appendix A can be summarized as following.
For $J'<J'_c$ and magnetic fields $h>h_c=\Delta$, $m_s^{\perp}\propto 
\sqrt{(h-h_c)}$ is expected, consistent with bond-operator mean-field theory~\cite{matsumoto_pressure}. 
At the critical inter-dimer coupling, $J'=J'_c$, a linear relation $m_s^{\perp}\propto h$ is expected.
Within the antiferromagnetically ordered regime, $J'>J'_c$, a finite staggered magnetization $m_s^{\perp}=m_s$ exists.
For small fields, the order parameter scales with $m_s^{\perp}-m_s\propto h^2$, as presented in Ref.~\onlinecite{matsumoto_pressure}.

\begin{figure}[h]
\includegraphics[width=8cm]{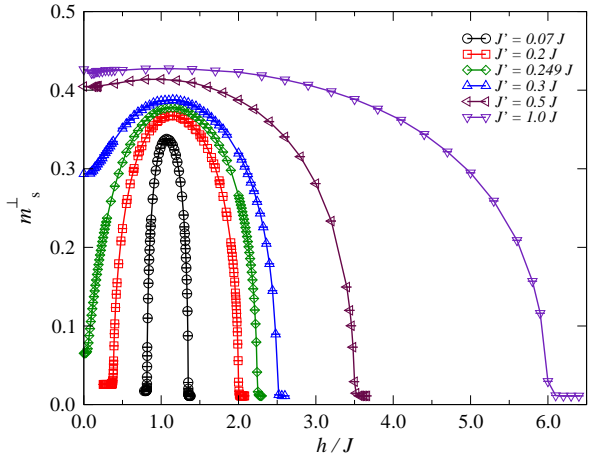}
\vspace{-0.3cm}
\caption{\label{smag} 
Zero-temperature staggered transverse magnetization $m_s^{\perp}$
in coupled aligned dimer arrays as a function of an applied 
magnetic 
field $h$. Quantum Monte Carlo data of a $16\times 16\times 16$ system are shown for different values of the inter-dimer coupling $J'$. The 
intra-dimer coupling is denoted by $J$. 
The off-sets of  $m_s^{\perp}$ outside the range $h_c\le h\le h_s$ 
are due to finite-size effects.
}
\end{figure}

QMC calculations of $m_s^{\perp}$ for different values of $J'/J$ in a system with $L=16$ are shown in Fig.~\ref{smag}.
This Figure demonstrates that for $J'\le J'_c$, the staggered magnetization
is largest for $h=(h_c+h_s)/2$ and decreases upon approaching $h_c$ and $h_s$. In the thermodynamic limit, $m_s^{\perp}$ vanishes in the fully polarized regime, $h>h_s$ and below $h_c$ for weakly coupled dimers.
In contrast to the uniform magnetization, the field dependence of $m_s^{\perp}$ shows a strong finite-size dependence.
Finite-size effects are most pronounced for magnetic field regimes outside the partially polarized phase, i.e. for $h<h_c$ and 
$h>h_s$, where
$m_s^{\perp}=0$ in the thermodynamic limit. On the finite systems accessible to QMC, the values of $m_s^{\perp}$, defined in Eq. (\ref{perp_stagg_mag}) do not vanish. However, they scale to zero upon increasing the system size.
In order to extract the scaling behavior of $m_s^{\perp}$, we thus need to perform a 
careful finite-size scaling analysis of the numerical data.

\begin{figure}[h]
\includegraphics[width=8.5cm]{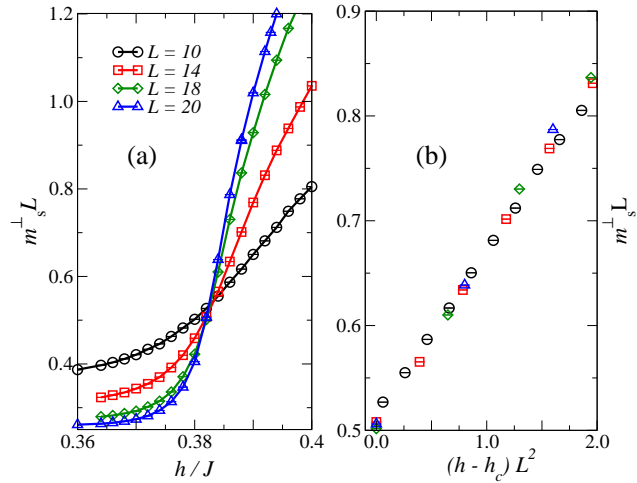}
\vspace{-0.3cm}
\caption{\label{smag_02} 
Scaling plot (a) and data collapse (b) for the zero-temperature staggered transverse magnetization $m_s^{\perp}$
of weakly coupled aligned dimers as a function of the applied magnetic field $h$ for a coupling-ratio of $J'/J=0.2$.
Results from quantum Monte Carlo simulations are shown for systems sizes of $L=10$ to $20$.
}
\end{figure}

Such an analysis can be directly applied in the weak coupling regime, for $J'<J'_c$. Here, the field-induced ordering 
transition is known 
to constitute a Bose-Einstein condensation of triplons,\cite{sachdevbook}
with a dynamical critical exponent $z=2$, reflecting the 
quadratic dispersion relation of these bosonic excitations.
Therefore, the quantum phase transition at the critical field $h_c$ is in the universality class of the classical 5-dimensional $O(2)$ model, and a finite-size scaling analysis similar to the one performed for the 
zero-field quantum critical point in Sec. \ref{mu_sec} can be applied. Since the classical theory is now above $d_c=4$, mean-field scaling without logarithmic corrections applies.
In particular, at the critical field $h_c$, one finds
\be
\label{m_l_scale}
m_{\rm s}^{\perp} \; L \propto F(L/\xi)\;,
\ee
where $F$ is a scaling function that depends on the ratio of the correlation length $\xi$ and the system size $L$.
In order to determine $h_c$ for a given value of $J'<J_c$, we can thus construct a scaling plot of $m_s^{\perp} L$ as a function of $h/J$.
Fig.~\ref{smag_02} illustrates an example of such a plot for
an inter-dimer coupling $J'=0.2 J$. In Fig.~\ref{smag_02}(a), 
data for $m_s^{\perp} L$ are shown for system sizes $L$ = 10, 14, 18, and 20.
The intersection point of the different finite-size data allows the extraction of the critical field $h_c=0.383\pm 0.001$ for this value of 
$J'=0.2 J$.
 
Once the critical field $h_c$ is determined, data collapse of the finite-size data
is verified in the vicinity of $h_c$.
Namely, from the scaling of the correlation length close to the critical field, $\xi\propto |h-h_c|^{- \nu}$, one obtains
\be
\label{data_collapse}
m_{\rm s}^{\perp} \; L \propto \tilde{F}(L^{1/\nu}\; |h-h_c|\;)\;,
\ee
with a new scaling function $\tilde{F}$ and $\nu=1/2$.
An example of the data collapse for $J'=0.2 J$ is shown in Fig.~\ref{smag_02}~(b).

\begin{figure}[h]
\includegraphics[width=8.5cm,height=8cm]{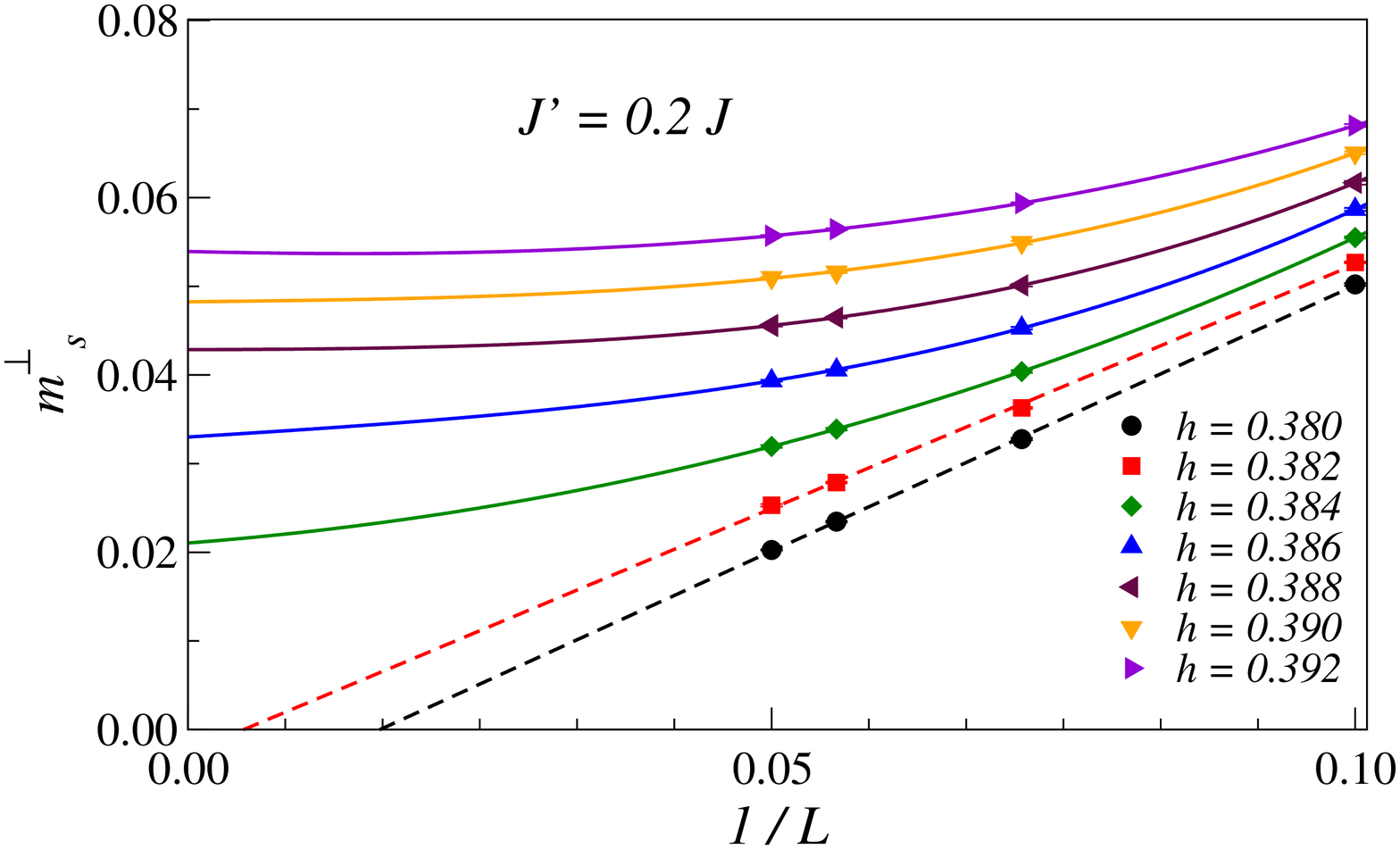}
\vspace{-2.0cm}
\caption{\label{smag_ex_02}
Extrapolation of finite-size data of staggered magnetization for different
magnetic fields close to the transition. At $h_c$, thermodynamic limit is obtained
from a $1/L$-extrapolation. Higher order polynomials are required away from
criticality.
For all data presented here, up to the third order is used.
Dashed lines are for $h<h_c$.
}
\end{figure}

\begin{figure}[h]
\includegraphics[width=8.0cm]{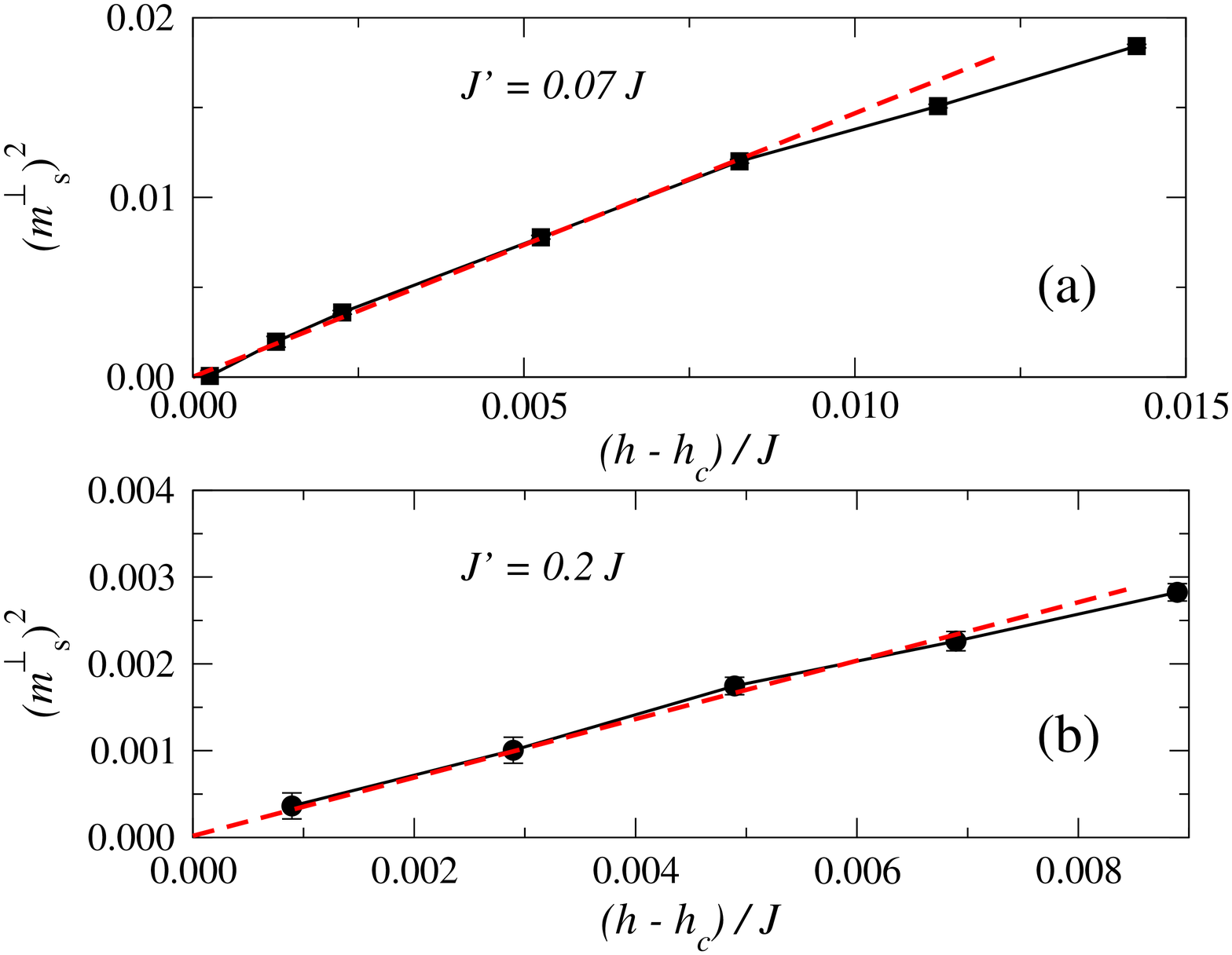}
\includegraphics[width=8.0cm]{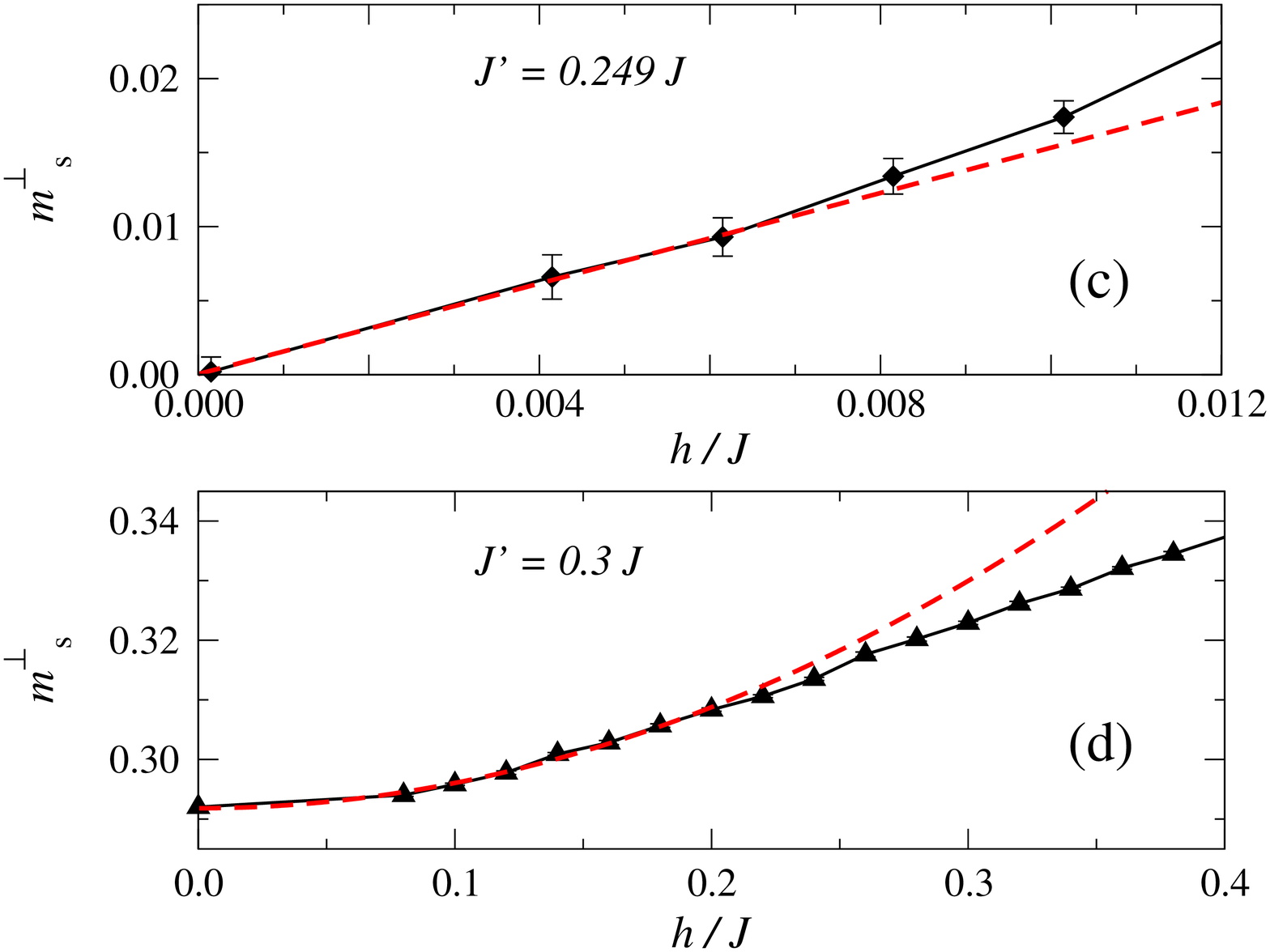}
\vspace{-0.1cm}
\caption{\label{m_h_inf} 
Scaling behavior of the zero-temperature staggered transverse magnetization
as a function of the applied magnetic field $h$. Quantum Monte Carlo data
are shown after extrapolation of finite-size data ($L=10-20$) to the thermodynamic limit.
(a)(b) in the
dimer spin liquid phase ($J' < J'_c$), $(m_{\rm s}^{\perp})^2$ scales linear
with $h - h_c$.
(c) near the critical coupling ($J'_c$), 
$m_{\rm s}^{\perp}$ scales linear with $h - h_c$.
(d) for $J' > J'_c$, $m_{\rm s}^{\perp}-m_s$ increases quadratically with $h$.
} 
\end{figure}

In order to extract the behavior of $m_s^{\perp}(h)$ for $h>h_c$ in the thermodynamic limit,
we perform an extrapolation of the finite-size data $m_s^{\perp}(L)$ as a function
of $1/L$. 
An example of such an extrapolation, again for $J'/J=0.2$ is shown in Fig.~\ref{smag_ex_02} for various values of $h$.
While for $h$ close to $h_c$, a linear scaling in $1/L$ is obtained, extrapolation away from $h_c$ requires
the use of higher order polynomials. Using these extrapolated values of $m_s^{\perp}(h)$, representing the thermodynamic limit, 
we then obtained the scaling behavior of $m_s^{\perp}(h)$ shown for different values of $J'/J$ in Fig.~\ref{m_h_inf}.

For weakly coupled dimers with $J'=0.07J$ Fig.~\ref{m_h_inf}(a) exhibits a scaling of $m_{\rm s}^{\perp}\propto \sqrt{h-h_c}$.
The scaling exponent $\frac{1}{2}$ is also found for stronger
inter-dimer couplings, as shown in Fig.~\ref{m_h_inf}(b) for $J'=0.2J$.
This square root dependence $m_{\rm s}^{\perp} \propto \sqrt{h - h_{c}}$ 
is consistent with the analytical results and expected from the mean-field value $\beta=1/2$.

Next, let us consider the scaling of $m_s^{\perp}(h)$ for a critical inter-dimer coupling $J'=J'_c$. In Fig.~\ref{m_h_inf}(c),
we show results from simulations at the critical point $J'=0.249$. 
Here, a linear scaling
$m_{\rm s}^{\perp} \propto h $ is observed, consistent with the Ginzburg-Landau calculations from Appendix A.
Finally, we consider the N\'eel-ordered phase  ($J'>J'_c$). The results from the extrapolated QMC data is shown for $J'=0.3 J$ in 
Fig.~\ref{m_h_inf}(d).  Increasing the magnetic field, a non-linear increase of $m_s^{\perp}(h)$ from its zero-field value $m_s$ is found, 
which can
be fitted well to the analytical prediction, $m_s^{\perp}(h)-m_s\propto h^2$.
We thus obtain agreement of the QMC data with the analytical results based on Ginzburg-Landau and bond-operator mean-field 
theory also for the scaling of the order parameter.
\vspace{-2mm}

%\vspace{-1cm}
\section{\label{sum_sec}Summary}
We  examined quantum phase transitions in three-dimensional coupled dimer arrays.
The zero-temperature phase diagrams of these systems feature a low-field dimer spin liquid phase at  weak inter-dimer couplings, a partially polarized regime with 
long-range transverse magnetic order for intermediate magnetic fields
$h_c\le h\le h_s$, and a fully polarized phase at high magnetic fields.

The critical exponents associated with quantum phase transitions
between these regimes were extracted using
finite-size scaling analysis of quantum Monte Carlo data. 
The numerical values of these exponents compare well with 
Ginzburg-Landau calculations and bond-operator mean-field theory. 
In particular, for small inter-dimer coupling, the order parameter is found to scale 
as $(h - h_{c})^{1/2}$, whereas the uniform magnetization scales linear in $h - h_{c}$. Moreover, at the quantum critical point $(J'/J)_c$, we demonstrated linear scaling for $m_{\rm s}^{\perp} \propto h$.
In a recent magnetization study on the $\rm TlCuCl_3$ system
under hydrostatic pressure~\cite{tanaka_pressure2},
a cubic scaling of the uniform magnetization $m_u\propto h^3$ at the 
critical value of the applied pressure was observed. This
is in perfect agreement with numerical results and analytical predictions based on Ginzburg-Landau calculations and bond-operator theory.

We found that corrections to mean-field scaling emerge at the 
zero-field pressure-induced quantum phase 
transition, such as those in
Ref.~\onlinecite{tanaka_pressure,tanaka_pressure2}. 
For example, in 
Ref. \onlinecite{tanaka_pressure2}, the pressure-dependence of the gap was fitted to a power law with an exponent 0.445, while 
the mean-field theory predicts an exponent 0.5, without taking logarithmic corrections into account.\cite{matsumoto_pressure}
Assuming a linear scaling of the pressure $P$ with the inter-dimer interactions,~\cite{tanaka_pressure2} close to the critical point $P_c$,
we find the scaling of the 
gap $\Delta$ to be
\be
\Delta \propto 1/\xi \propto 
(P-P_c)^{1/2}\left(-\ln\left|\frac{P-P_c}{P_c}\right|\right)^{1/6}.
\ee
from a renormalization group analysis of the classical $\phi^4$-theory at 
$d_c=4$~\cite{zinn_justin}.
Therefore, careful fitting of experimental data can exhibit the logarithmic 
corrections near quantum critical points in three-dimensional compounds.

\section*{Acknowledgment}
We thank Andreas Honecker,
Bruce Normand, Tommaso Roscilde, Subir Sachdev
Manfred Sigrist, Hidekazu Tanaka, Matthias Troyer, and Matthias Vojta for useful  discussions. 
Furthermore, we acknowledge financial support from NSF Grant No. 
DMR-0089882. Computational support was provided 
by the USC Center for High Performance Computing and Communications.
Parts of the numerical simulations were performed using the ALPS project library.~\cite{ALPS}

\appendix

\section{\label{gb_sec}Ginzburg-Landau Theory}
In this Appendix, we present a phenomenological 
Ginzburg-Landau description of weakly coupled dimers in a magnetic field~\cite{ehrenfest}, 
and derive the scaling exponents of the uniform magnetization $m_u$ and
the order parameter $m_s^{\perp}$ in the partially polarized phase.

Due to rotational symmetry
within the plane perpendicular to the external field, the free
energy of the system can be expanded as
\begin{equation}
\label{free_e}
F = a(h,J') \; (m_{\rm s}^{\perp})^2 + b \; (m_{\rm s}^{\perp})^4 ,
\end{equation}
neglecting higher orders of $m_{\rm s}^{\perp}$
in the vicinity of the critical field.\cite{GL}
Here, $a(h,J')$ vanishes for $h$ equal to the critical magnetic field $h=h_c(J')$.
Taking into account time reversal
symmetry, $a$ can be expressed as
$a = \tilde{ a}\cdot({h}^2 -h_c^2(J'))$.
The staggered magnetization
$m_{\rm s}^{\perp}$ and the uniform magnetization $m_u$ is obtained
from Eq.~\ref{free_e} as
\be
\frac{\partial F}{\partial m_{\rm s}^{\perp}} \Bigg|_{h,J'}= 0 \;\; 
\mbox{and} \;\;\; m_u = -\frac{\partial F}{\partial h} \Bigg|_{J'}.
\ee
From this, one finds
\begin{eqnarray}
\label{gl_ms_mu}
m_{\rm s}^{\perp} &=& \sqrt{\frac{\tilde{a}(h^2-h_c^2)}{2b}} \\
m_u &=& 2 \tilde{a} h (m_{\rm s}^{\perp})^2\;.
\end{eqnarray}

To proceed further, we distinguish two regions, namely weak inter-dimer
couplings ($J'<J'_c$) and the quantum critical point ($J' = J'_c$). For the
first region, (\ref{gl_ms_mu}) can be expanded in terms of the distance to the
lower critical field, i.e. $h-h_{c}$. One obtains
\begin{eqnarray}
\hspace{-2cm} \mbox{for } J' < J'_c\; \mbox{:} \hspace{0.5cm}&&
\hspace{2cm} \nonumber \\ 
m_{\rm s}^{\perp} &=& \sqrt{
\frac{\tilde{a}(h+h_{c})(h-h_{c})}{2b}}, \nonumber \\
\Rightarrow m_{\rm s}^{\perp} &\propto& (h-h_{c})^{1/2}\;,\\
\Rightarrow m_u &\propto& 2 \tilde{a} h (h-h_{c})^{1}\;.
\end{eqnarray}

At the critical inter-dimer coupling the critical field is zero. 
Therefore,
\begin{eqnarray}
\hspace{-2cm} \mbox{for } J' = J'_c\; \mbox{:} \hspace{0.8cm}&&
\hspace{3cm} \nonumber \\ 
m_{\rm s}^{\perp} &=& \sqrt{
\frac{\tilde{a}}{2b}} \; \; h\;, \\
\Rightarrow m_u &\propto& 2 \tilde{a} h^{3} .
\end{eqnarray}

These exponents agree with the numerical data 
presented in the preceding sections, as well as 
previous bond-operator mean-field theory results~\cite{matsumoto_pressure}.

\end{document}